\documentclass{llncs}
\usepackage[pdftex]{graphicx}
\graphicspath{./sudoku.png}
\graphicspath{./PGAS.png}
\graphicspath{./ClassDiagramX10.png}
\graphicspath{./big-speed-ups.png}

\usepackage{algorithm}
\usepackage{algpseudocode}

\usepackage{listings}
\newcommand{\BL}{\vspace{\baselineskip}}

\makeatletter
\renewcommand{\ALG@beginalgorithmic}{\small}
\makeatother

\begin{document}
\title{Experimenting with X10 for Parallel Constraint-Based Local Search}

\author{Danny Munera\inst{1} \and Daniel Diaz\inst{1} \and Salvador Abreu\inst{2}}

\institute{
University of Paris 1-Sorbonne, France \\
\email{\{Danny.Munera,Daniel.Diaz\}@univ-paris1.fr}
\and
Universidade de \'Evora and CENTRIA, Portugal \\
\email{spa@di.uevora.pt}}

\maketitle

\begin{abstract}
  In this study, we have investigated the adequacy of the PGAS
  parallel language X10 to implement a Constraint-Based Local Search
  solver.  We decided to code in this language to benefit from the
  ease of use and architectural independence from parallel resources
  which it offers.  We present the implementation strategy, in search
  of different sources of parallelism in the context of an
  implementation of the Adaptive Search algorithm.  We extensively
  discuss the algorithm and its implementation.  The performance
  evaluation on a representative set of benchmarks shows close to
  linear speed-ups, in all the problems treated.
\end{abstract}

\section{Introduction}

Constraint Programming has been successfully used to model and solve
many real-life problems in diverse areas such as planning, resource
allocation, scheduling and product line
modeling~\cite{Rossi2006,Salinesi}.  Classically constraint
satisfaction problems (CSPs) may be solved exhaustively by complete
methods which are able to find all solutions, and therefore determine
whether any solutions exist.  However efficient these solvers may be,
a significant class of problems remains out of reach because of
exponential growth of search space, which must be exhaustively
explored. 
Another approach to solving CSPs entails giving up completeness and
resorting to (meta-) heuristics which will guide the process of
searching for solutions to the problem.  Solvers in this class make
choices which limit the part of the search space which actually gets
visited, enough so to make problems tractable.  For instance a
complete solver for the \emph{magic squares} benchmark will fail for
problems larger than $15\times15$ whereas a local search method will
easily solve a $100\times100$ problem instance within the lower
resource bounds.  On the other hand, a local search procedure may not
be able to find a solution, even when one exists.

However, it is unquestionable that the more computational resources
are available, the more complex the problems that may be solved.  We
would therefore like to be able to tap into the forms of augmented
computational power which are actually available, as conveniently as
feasible.  This requires taming various forms of explicitly parallel
architectures.

Present-day parallel computational resources include increasingly
multi-core processors, General Purpose Graphic Processing Units
(GPGPUs), computer clusters and grid computing platforms.  Each of
these forms requires a different programming model and the use of
specific software tools, the combination of which makes software
development even more difficult.

The foremost software platforms used for parallel programming include
POSIX Threads~\cite{Butenhof1997} and OpenMP~\cite{OpenMP} for
shared-memory multiprocessors and multicore CPUs, MPI~\cite{Snir1996}
for distributed-memory clusters or CUDA~\cite{NVIDIAa} and
OpenCL~\cite{Opencl} for massively parallel architectures such as
GPGPUs.  This diversity is a challenge from the programming language
design standpoint, and a few proposals have emerged that try to
simultaneously address the multiplicity of parallel computational
architectures.


Several modern language designs are built around the Partitioned
Global Address Space (PGAS) memory model, as is the case with
X10~\cite{Saraswat2012}, Unified Parallel C~\cite{El-Ghazawi2005} or
Chapel~\cite{CrayInc.2012}.  Many of these languages propose
abstractions which capture the several forms in which multiprocessors
can be organized.  Other, less radical, approaches consist in
supplying a library of inter-process communication which relies on and
uses a PGAS model.


In our quest to find a scalable and architecture-independent
implementation platform for our exploration of high-performance
parallel constraint-based local search methods, we decided to
experiment with one of the most promising new-generation languages,
X10~\cite{Saraswat2012}.

The remainder of this article is organized as follows: Section~\ref{sec:CBLS}
present some necessary background in Constraint-Based Local Search.
Section~\ref{sec:PGASX10} discusses the PGAS Model and presents X10 language.
The sources of parallelism in the Adaptive Search algorithm and its X10
implementation are the object of Section~\ref{sec:X10parallel}.
Section~\ref{sec:eval} describes the benchmarks used in our experiments.  The
results and a discussion of the performance evaluation may be found in
Section~\ref{sec:PerformanceAnalysis}. A short conclusion ends the paper.


\section{Constraint-Based Local Search}
\label{sec:CBLS}
It is possible to classify the method for constraint solving in two
groups: \emph{complete} or \emph{incomplete} methods. Complete methods
always find a solution of the problem, if any exists. There are two
main groups of algorithms: search and inference. Firstly, a
representative complete search method is depth-first backtracking,
which incrementally builds a solution exploring as far as possible
along each branch of the search space, such that no constraint is
violated. If from some point no value allows to extend the current partial
assignment, it backtracks and a previous assignment is
reconsidered. Secondly, the inference methods use constraints to
reduce the number of legal values for a variable, and propagate this
new domain to reduce other domain variables, finding a solution.

Incomplete methods, like Local Search, try to found a solution using
limited resources, however this type of methods does not guarantee
neither to find one solution nor to detect an inconsistency problem
(no possible solution). Local search is suitable for optimization
problems with a cost function which makes possible to evaluate the
quality of a given configuration (assignment of variables to current
values). It also needs a transition function which defines, for each
configuration, a set of neighbors. The simplest Local Search algorithm
start from a random configuration, explores the neighborhood and then
selects a promising neighbor and moves there. This is a iteratively
process that continues until a solution is found.

In this study, a Local Search method, the Adaptive Search algorithm is
selected. The Adaptive Search~\cite{Codognet2001,Codognet2003}
is a generic, domain-independent constraint-based local search
method. This meta-heuristic takes advantage of the CSP
formulation and makes it possible to structure the problem in terms of
variables and constraints and to analyze the current assignment of
variables more precisely than an optimization of a global cost
function e.g. the number of constraints that are not
satisfied. Adaptive Search also includes an adaptive memory inspired
in Tabu Search~\cite{Glover1997} in which each variable leading to a
local minimum is marked and cannot be chosen for the next few
iterations. A local minimum is a configuration for which none of the
neighbors improve the current configuration.  The input of the
Adaptive Search algorithm is a CSP, for each constraint an
error function is defined. This function is a heuristic value to
represent the degree of satisfaction of a constraint and gives an
indication on how much the constraint is violated.

Adaptive Search is based on iterative repair from the variables and
constraint error information, trying to reduce the error in the worse
variable. The basic idea is to calculate the error function for each
constraint, and then combine for each variable the errors of all
constraints in which it appears, thus projecting constraint errors on
involved variables. Then, the algorithm chooses the variable with the
maximum error as a ``culprit'' and selects it to modify later its
value.

The purpose is to select the best neighbor move for the culprit
variable, this is done by considering all possible changes in the value
of this variable (neighbors) and selecting the lower value of the
overall cost function. Finally, the algorithm also includes partial
resets in order to escape stagnation around local minima; and it is
possible to restart from scratch when the number of iterations becomes
too large.
Algorithm~\ref{fig:as-algo} presents a particular implementation of
the Adaptive Search algorithm dedicated to permutation problems. In
this case all $N$ variables have the same initial domain of size $N$ and
are subject to an implicit \emph{all-different} constraint.


\section{PGAS model and X10}
\label{sec:PGASX10}

The current arrangement of tools to exploit parallelism in machines
are strongly linked to the platform used. As it was said above, two
broad programming models stand out in this matter: \emph{distributed}
and \emph{shared memory} models. For large distributed memory systems,
like clusters and grid computing, Message Passing Interface
(MPI)~\cite{Snir1996} is a de-facto programming standard. The
key idea in MPI is to decompose the computation over a collection of
processes with private memory space. This processes can communicate
with each other through message passing, generally over a
communication network.

With the recent growth of many-core architectures, the shared memory
approach have increased its popularity. This model decomposes the
computation in multiple threads of execution sharing a common address
space, communicating with each other by reading and writing shared
variables. Actually, this is the model used by traditional programming
tools like Fortran or C through libraries like
\emph{pthreads}~\cite{Butenhof1997} or OpenMP~\cite{OpenMP}.

\begin{figure}[h]
\centerline{\fbox{{\includegraphics[scale=0.37]{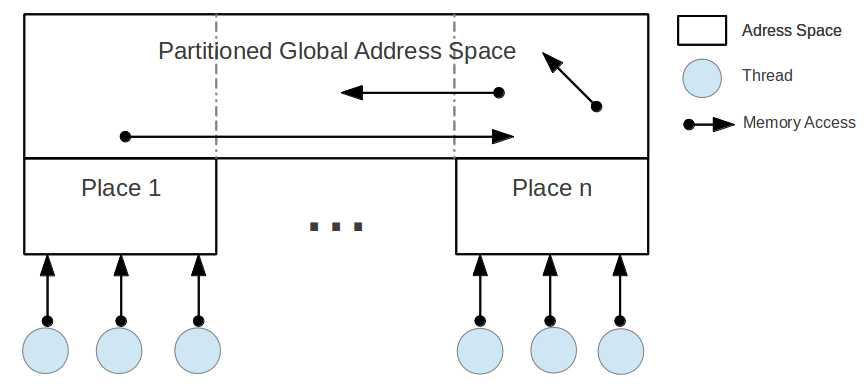}}}} 
\caption{PGAS Model}\label{fig:PGASModel}
\end{figure}

The PGAS model tries to combine the advantages of the two approaches
mentioned so far.  This model extends shared memory to a distributed
memory setting.  The execution model allows having multiple processes
(like MPI), multiple threads in a process (like OpenMP), or a
combination (see Figure~\ref{fig:PGASModel}).  Ideally, the user would
be allowed to decide how tasks get mapped to physical resources.
X10~\cite{Saraswat2012}, Unified Parallel C~\cite{UPCConsortium2005}
and Chapel~\cite{CrayInc.2012} are examples of PGAS-enabled languages,
but there exist also PGAS-based IPC libraries such as
GPI~\cite{Machado2009}, for use in traditional programming languages.
For the experiments described herein, we used the X10 language.

\subsection{X10 in a Nutshell}
\label{sec:X10}
X10~\cite{Saraswat2012} is a general-purpose language developed by
IBM, which provides a PGAS variation: Asynchronous PGAS (APGAS). APGAS
extends the PGAS model making it flexible, even in non-HPC
platforms~\cite{Saraswat2010}. Through this model X10 can support
different levels of concurrency with simple language constructs.

There are two main abstractions in X10 model: \emph{places} and
\emph{activities}. A \emph{place} is the abstraction of a virtual
shared-memory process, it has a coherent portion of the address space
together with threads (activities) that operate on that memory. The
X10 construct for creating a place in X10 is \emph{at}, and is commonly
used to create a place for each processing unit in the platform. An
\emph{activity} is the mechanism to abstract the single threads that
perform computation within a place. Multiple activities may be active
simultaneously in a place.

With these two components of X10 implement the main concepts of the
PGAS model.  Howevere, the language includes other interesting tools
with the goal of improving the abstraction level of the language.
Synchronization is supported thanks to various operations such as
\emph{finish}, \emph{atomic} and \emph{clock}. The operation
\emph{finish} is used to wait the termination of a set of activities,
it behaves like a traditional barrier. The constructs \emph{atomic}
ensures an exclusive access to a critical portion of code. Finally,
the construct \emph{clock} is the standard way to ensure the
synchronization between activities or places.  X10 supports the
distributed array construct, which makes it possible to divide an
array into sub-arrays which are mapped to available places. Doing this
ensures a local access from each place to the related assigned
sub-array. A detailed examination of X10, including tutorial, language
specification and examples can be consulted at
\url{http://x10-lang.org}.


\section{ X10 Adaptive Search Parallel Implementation}
\label{sec:X10parallel}

In order to get advantage of the parallelism it is necessary to identify
the sources of parallelism of the algorithm.  In~\cite{Crainic2010},
the authors survey the state-of-the-art of the main parallel
meta-heuristic strategies and discuss general design and
implementation principles.  This study raises a number of important
issues in the taxonomy of parallel sources that lies in
meta-heuristics. They classify the decomposition of activities for
parallel work in two main groups: \emph{functional parallelism} and
\emph{data parallelism}.

On the one hand, in \emph{functional parallelism} generally different
tasks work in parallel on the same data, allocated in different
compute instances. On the other hand, \emph{data parallelism} refers
to the methods in which the problem domain or the associated search
space is decomposed. A particular solution methodology is used to
address the problem on each of the resulting components of the search
space.  Based on this study, we explored the parallelism in the
Adaptive Search method in both \emph{functional parallelism} and
\emph{data parallelism} and this article mostly reports on the latter,
because this is where we expect the most significant gains to show.

\subsection{Adaptive Search X10 sequential implementation}

The first stage of the X10 implementation was to develop a sequential
algorithm, which is described as algorithm~\ref{fig:as-algo}.

\begin{figure}[htb]
\centerline{\fbox{{\includegraphics[scale=0.40]{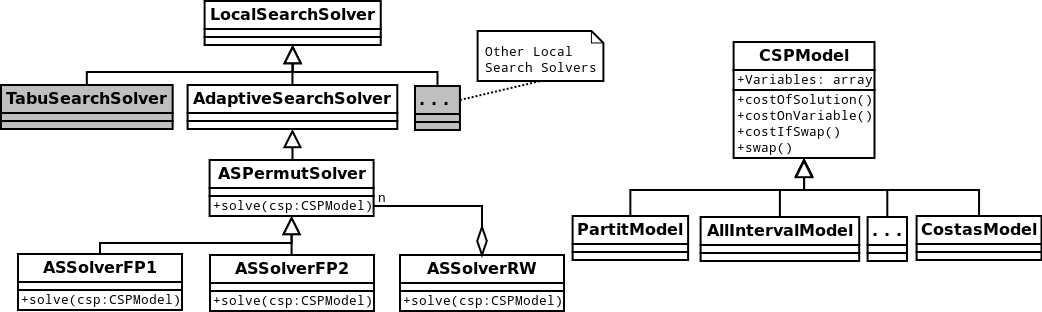}}}} 
\caption{X10 Class Diagram basic project}\label{fig:ClassDiagramX10}
\end{figure}

Figure~\ref{fig:ClassDiagramX10} shows the class diagram of the basic
X10 project. The class \emph{ASPermutSolver} contains the Adaptive
Search permutation specialized method implementation. This class
inherits the basic functionality from a general implementation of the
Adaptive Search solver (in class \emph{AdaptiveSearchSolver}), which
in turn inherits a very simple Local Search method implementation from
the class \emph{LocalSearchSolver}.  This class is then specialized
for different parallel approaches, which we experimented
with.\footnote{We experimented with two versions of Functional
  Parallism (FP1 and FP2) and a Random Walk version (RW.)}

\begin{algorithm}[H]
   \caption{Adaptive Search Base Algorithm}
   \label{fig:as-algo}
     \BL
{\textbf{Input}: problem given in CSP format:}

\begin{itemize}
\item  set of variables $V=\{X_1,X_2\cdots\}$ with their domains
\item  set of constraints $C_j$ with error functions
\item  function to project constraint errors on vars (positive) cost function to minimize
\item  $T$: Tabu tenure (number of iterations a variable is frozen on local minima)
\item  $RL$: number of frozen variables triggering a reset
\item  $MI$: maximal number of iterations before restart
\item  $MR$: maximal number of restarts
\end{itemize}

{\textbf{Output}}: a solution if the CSP is satisfied or a quasi-solution of minimal cost otherwise.

     \BL

   \begin{algorithmic}[1]
     \State $Restart \gets 0$
     \Repeat
       \State $Restart \gets Restart + 1$
       \State $Iteration \gets 0$
       \State Compute a random assignment $A$ of variables in $V$
       \State $Opt\_Sol \gets A$
       \State $Opt\_Cost \gets cost(A)$
       \Repeat
         \State $Iteration \gets Iteration +1$
         \State Compute errors constraints in $C$ and project on relevant variables
         \State Select variable $X$ with highest error: $MaxV$
         \State \Comment{not marked Tabu}
         \State Select the move with best cost from $X$: $MinConflictV$ 
         \If{no improvement move exists}
           \State mark $X$ as Tabu for $T$ iterations
           \If{number of variables marked Tabu $\geq RL$}
             \State randomly reset some variables in $V$
              \State \Comment{and unmark those Tabu}
           \EndIf
         \Else
           \State swap($MaxV$,$MinConflictV$),
           \State \Comment{modifying the  configuration $A$}
           \If{$cost(A) < Opt\_Cost$}
             \State $Opt\_Sol \gets A$
             \State $Opt\_Cost \gets costs(A)$
           \EndIf
         \EndIf
       \Until{$Opt\_Cost = 0$ (solution found) or $Iteration \geq MI$}
     \Until{$Opt\_Cost = 0$ (solution found) or $Restart \geq MR$}
     \State $output (Opt\_Sol, Opt\_Cost)$
   \end{algorithmic}
 \end{algorithm}

Moreover,
a simple CSP model is described in the class \emph{CSPModel}, and
specialized implementations of each CSP benchmark problem are
contained in the classes \emph{PartitModel, MagicSquareModel,
  AllIntervallModel} and \emph{CostasModel}, which have all data
structures and methods to implement the error function of each
problem.

Listing~\ref{fig:as-x10} shows a simplified skeleton code of our X10
sequential implementation, based on Algorithm~\ref{fig:as-algo}.  The
core of the Adaptive Search algorithm is implemented in the method
solve. The \emph{solver} method receives a \emph{CSPModel} instance as
a parameter. In line~\ref{as-x10-decl1}, the CSP variables of the
model are initialized with a random permutation, in the next line the
total cost of the current configuration is computed. The sentence
\emph{while} on line~\ref{as-x10-decl2} begins the main loop of
the algorithm. The \emph{selectVarHighCost} function
(Line~\ref{as-x10-decl3}) selects the variable with the maximum error
and saves the result in \emph{maxI} variable. The
\emph{selectVarMinConflict} function (Line~\ref{as-x10-decl4}) selects
the best neighbor move from the high cost variable \emph{maxI}, and
saves the result index in \emph{minJ} variable. Finally, if no local
minimum is detected, the algorithm swaps the variables \emph{maxI} and
\emph{minJ} (permutation problem) and recompute the total cost of the
current configuration (Line~\ref{as-x10-decl5}). The solver function
ends if the \emph{totalCost} variable reaches the $0$ value or if the
algorithm reaches the maximum number of iterations.

\begin{lstlisting}[caption={Simplified AS X10 Sequential Implementation},label=fig:as-x10]
class ASPermutSolver {
  var totalCost: Int;
  var maxI: Int;
  var minJ: Int;

  public def solve (csp: CSPModel): Int {
    // (...local variables...)
(*\label{as-x10-decl1}*)    csp.initialize();
    totalCost = csp.costOfSolution();
(*\label{as-x10-decl2}*)    while (totalCost != 0) {
      // (...restart code...)
(*\label{as-x10-decl3}*)      maxI = selectVarHighCost (csp);
(*\label{as-x10-decl4}*)      minJ = selectVarMinConflict (csp);
      // (...local min tabu list, reset code...)
      csp.swapVariables (maxI, minJ);
(*\label{as-x10-decl5}*)      totalCost = csp.costOfSolution ();
    }
    return totalCost;
  }
}
\end{lstlisting}

\ifx{1}{2}                      
\subsection{Adaptive Search X10 functional parallelism implementation}
Functional parallelism is our first attempt to parallelize
the Adaptive Search algorithm. The key aim for this
implementation is to decompose the problem into different tasks, each
task works in parallel on the same data. To achieve this objective is
necessary to change the inner loop of the sequential Adaptive Search
algorithm.

In this experiment, we decided to change the structure of the function
\emph{selectVarHighCost}, because therein lies the most costly activities
performed in the inner loop. As it was described in Section~\ref{sec:CBLS},
the most important task performed by this function is to go through the
variables array of the CSP model, to compute the cost of each variable and to
select the variable with the highest cost.  A
X10 skeleton implementation of \emph{selectVarHighCost} function is presented
in Listing~\ref{fig:functSelVarHighCost-x10}.

\begin{lstlisting}[caption={Function selVarHighCost in X10},label=fig:functSelVarHighCost-x10]
public def selectVarHighCost( csp : ModelAS ) : Int {
	// (...local variables...)
	// main loop: go through each variable in the CSP
	for (i = 0; i < size; i++) {
		// (...count marked variables...)
		cost = csp.costOnVariable (i);
		// (...select the highest cost...)
	}
	return maxI;	// (index of the highest cost)
}
\end{lstlisting}

It is worth noticing that if the problem has a great number of variables,
this function must process the entire variables vector at each iteration. It
is then natural to try to parallelize this task.  We developed a \emph{first
  approach}, in which $n$ single activities are created at each iteration.
Each activity processes a portion of the variables array and performs the
required computations.  The X10 construct \emph{async} was selected to create
individual \emph{activities} sharing the global
array. Listing~\ref{fig:as-x10FP1} shows the X10 skeleton code for the
\emph{first approach} of the \emph{functional parallelism} in the function
\emph{selectVarHighCost}.

\begin{lstlisting}[caption={First approach to \emph{functional parallelism}},label=fig:as-x10FP1]
public def selectVarHighCost (csp : ModelAS) : Int {
//Initialization of Global variables
(*\label{x10FP1-decl1}*)  finish for(th in 1..THNUM){
(*\label{x10FP1-decl2}*)    async{
(*\label{x10FP1-decl3}*)      for (i = ((th-1)*partition); i < th*partition; i++){
        //(...calculate individual cost of each variable...)
        //(...Save variable with higher cost...)
     }
    }
  }
  //(...Terminate function: merge solutions...)
  return maxI; //(Index of the higher cost)
}
\end{lstlisting}

In this implementation the constant \emph{THNUM} in line~\ref{x10FP1-decl1}
represents the number of concurrent activities that are deployed by the
program; in the same line the keyword \emph{finish} ensures the termination
of all spawned activities; finally, the construct \emph{async} in
line~\ref{x10FP1-decl2} spawn independent individual tasks to cross over a
portion of the variables array (sentence \emph{for} in
line~\ref{x10FP1-decl3}).  With this strategy we face up with a well known
problem of functional parallelism: the overhead due to the management of
fine-grained activities. As expected results are not good enough (see
Section~\ref{sec:PerformanceAnalysis} for detailed results).

In order to limit the overhead due to activity creation, we implemented a
\emph{second approach}. Here $n$ working activities are created at the very
beginning of the solving process, just before the main loop of the algorithm.
These activities are available for all subsequent iterations.  However, it is
necessary to develop a synchronization mechanism to assign tasks to the
working activities and to wait for their termination. For this we created two
new classes: \emph{ComputePlace} and \emph{ActivityBarrier}.
\emph{ComputePlace} is a compute instance, which contains the functionality
of the working activities. \emph{ActivityBarrier} is a very simple barrier
developed with X10 monitors (X10 concurrent package).

Listing~\ref{fig:as-x10FP2} shows the X10 implementation of the
\emph{second approach}. 

\begin{lstlisting}[caption={Second approach to \emph{functional parallelism}},label=fig:as-x10FP2]
public class ASSolverFP1 extends ASPermutSolver{
(*\label{x10fp2-decl1}*)   val computeInst : Array[ComputePlace];
   var startBarrier : ActivityBarrier;
(*\label{x10fp2-decl2}*)   var doneBarrier : ActivityBarrier;

   public def solve(csp : ModelAS):Int{     
(*\label{x10fp2-decl3}*)      for(var th : Int = 1; th <= THNUM ; th++)
         computeInst(th)(*\footnote{Remark: in X10 the array notation is \textit{table(index)} instead of \textit{table[index]} as in C.}*) = new ComputePlace(th , csp);
      for(id in computeInst) 
(*\label{x10fp2-decl4}*)         async computeInst(id).run();
      
      while(total_cost!=0){
         // (...restart code...)
(*\label{x10fp2-decl5}*)         for(id in computeInst) 
(*\label{x10fp2-decl6}*)            computeInst(id).activityToDo = SELECVARHIGHCOST;
(*\label{x10fp2-decl7}*)         startBarrier.wait(); // send start signal
         // (...Activities working...) 
(*\label{x10fp2-decl8}*)         doneBarrier.wait(); // work ready
(*\label{x10fp2-decl9}*)         maxI=terminateSelVarHighCost();
         // (... local min tabu list, reset code ...)
      }
      //(Finish activities)
(*\label{x10fp2-decl10}*)      for(id in computeInst)
         computeInst(id).activityToDo = FINISH;
      startBarrier.wait();
      doneBarrier.wait();
(*\label{x10fp2-decl11}*)      return totalCost;
   }
}
\end{lstlisting}

This code begins with the definition of three global variables in
lines~\ref{x10fp2-decl1}-\ref{x10fp2-decl2}:~\emph{computeInst},
\emph{startBarrier} and \emph{doneBarrier}; \emph{computeInst} is an array of
\emph{ComputePlace} objects, one for each working activity
desired. \emph{startBarrier} and \emph{doneBarrier} are
\emph{ActivityBarrier} instances created to signalize the starting and ending
of the task in the compute place. In the
lines~\ref{x10fp2-decl3}-\ref{x10fp2-decl4}, before the main loop $THNUM$
working activities are created and started over an independent X10
activity. When the algorithm needs to execute the \emph{selectVarHighCost}
functionality, the main activity assigns this task putting a specific value
into the variable \emph{activityToDo} in the corresponding instance of the
\emph{ComputePlace} class (lines~\ref{x10fp2-decl5}~and~\ref{x10fp2-decl6}),
then the function \emph{wait()} is executed over the barrier
\emph{startBarrier} to notify all working activities to start
(line~\ref{x10fp2-decl7}). Finally, the function \emph{wait()} is executed
over the barrier \emph{doneBarrier} to wait the termination of the working
activities (line~\ref{x10fp2-decl8}), then on line~\ref{x10fp2-decl9} the
main activity can process the data with the function
\emph{terminateSelVarHighCost}. When the main loop ends all the working
activities are notified to end and the \textit{solve} function return
(lines~\ref{x10fp2-decl10}-\ref{x10fp2-decl11}).  Unfortunately, as we will
see below, the improvement of this \emph{second approach} is not important
enough (and in addition it has its own overhead due to synchronization
mechanisms).
\fi{}                           

\subsection{Adaptive Search X10 Parallel Implementation}
We first experimented with functional parallelism which consisted in
executing the inner loop in parallel, but the results were not
expressive, mostly because of thread handling overhead.  We then
implemented a data parallel variant, which turns out natural because
the sequential Adaptive Search algorithm can be used as an isolated
search instance.  Furthermore, the search space is divided using
different random start points (i.e configurations).  This strategy is
called Random Walks (RW) or Multi Search (MPSS, Multiple initial
Points, Same search Strategies)~\cite{Crainic2010} and has proven to
be very efficient~\cite{Diaz2011,Machado2012}.  The main point of our
study is to explore the adequacy of a programming language based on a
PGAS model: we will discuss the strengths and weaknesses of this
language when applied to Independent Random Walks, without any further
tuning of the parallel execution.

The implementation strategy is very different from the functional
parallelism: the key of this algorithm is to have several independent and
isolated instances of the Adaptive Search Solver applied to the problem
model. Then it is necessary to initialize the problem variables with a random
assignment of values for the variables and to distribute it to the available
processing resources in the computer platform. Finally, when one instance
reaches a solution, a termination detection communication strategy is used to
finalize the remaining running instances. This simple parallel version has no
inter-process communication, making it \emph{Embarrassingly} or
\emph{Pleasantly Parallel}.  The skeleton code of the algorithm is shown in
the Listing~\ref{fig:as-x10DP}.

\begin{lstlisting}[caption={Adaptive Search \emph{data parallel} X10 implementation},label=fig:as-x10DP]
public class ASSolverRW{
(*\label{x10dp-decl1}*)  val solDist : DistArray[ASPermutSolver];
(*\label{x10dp-decl2}*)  val cspDist : DistArray[CSPModel];
  def this( ){
    solDist=DistArray.make[ASPermutSolver](Dist.makeUnique());
    cspDist=DistArray.make[CSPModel](Dist.makeUnique());
  }	
  public def solve(){ 
    val random = new Random();
(*\label{x10dp-decl3}*)    finish for(p in Place.places()){ 
      val seed = random.nextLong();
(*\label{x10dp-decl4}*)      at(p) async {
(*\label{x10dp-decl5}*)        cspDist(here.id) = new CSPModel(seed);
(*\label{x10dp-decl6}*)        solDist(here.id) = new ASPermutSolver(seed);
(*\label{x10dp-decl7}*)        cost = solDist(here.id).solve(cspDist(here.id));
(*\label{x10dp-decl8}*)        if (cost==0){
          for (k in Place.places()) 
(*\label{x10dp-decl9}*)            if (here.id != k.id) 
(*\label{x10dp-decl10}*)              at(k) async{
                solDist(here.id).kill = true;
              }
(*\label{x10dp-decl11}*)        }
      }
    }
(*\label{x10dp-decl12}*)    return cost; 
}
\end{lstlisting}

For this implementation the \emph{ASSolverRW} class was created. This
class has two global distributed arrays \emph{solDist} and
\emph{cspDist} (lines~\ref{x10dp-decl1}~and~\ref{x10dp-decl2}). As
explained in Section~\ref{sec:X10}, the \emph{DistArray} class creates
an array which is spread across multiple X10 places.  In this case, an
instance of \emph{ASPermutSolver} and \emph{CSPModel} are spread over
all the available places in the program. On line~\ref{x10dp-decl3} a
finish operation is executed over a for loop that goes through all the
places in the program (\emph{Place.places()}). Then, an activity is
created in each place with the sentence \emph{at(p) async} on
line~\ref{x10dp-decl4}. Into the \emph{async} block, a new instance of
the solver (\emph{new ASPermutSolver(seed)}) and the problem (\emph{new
  CSPModel(seed)}) are created
(lines~\ref{x10dp-decl5}~and~\ref{x10dp-decl6}) and a random seed is
passed. In line~\ref{x10dp-decl7}, the solving process is executed and
the returned cost is assigned to the \textit{cost} variable. If this
cost is equal to $0$, the solver in a place has reached a valid
solution, it is then necessary to send a termination signal to the
remaining places (lines~\ref{x10dp-decl8}-~\ref{x10dp-decl11}). For
this, every place (i.e. every solver), checks the value of a
\textit{kill} variable at each iteration. When it becomes equal to
\textbf{true} the main loop of the solver is broken and the activity
is finished. To set a \textit{kill} remote variable from any X10 place
it was necessary to create a new activity into each remaining place
(sentence \emph{at(k) async} in line~\ref{x10dp-decl10}) and into the
\emph{async} block to change the value of the \textit{kill}
variable. Line~\ref{x10dp-decl9} with the sentence \emph{if (here.id
  != k.id)} filters all the places that not are the winner place
(\emph{here}). Finally, the function returns the solution of the
fastest place in line~\ref{x10dp-decl12}.


\section{Benchmark description}
\label{sec:eval}
In this study we used a set of benchmarks composed by four classical
problems in constraint programming: the magic square problem, the
number partitioning problem and the all-interval problem, all three
taken from the CSPLib~\cite{I.P.GentT.Walsh}; also we include the
Costas Array Problem (CAP) introduced in~\cite{Kadioglu2009}, which is
a very challenging real problem.

\subsubsection{Magic Square Problem (MSP)}
The Magic Square Problem (\texttt{prob019} in CSPLib) consists of
placing on a $N \times N$ square all the numbers in $\{1, 2, \ldots
,N^2\}$ such as the sum of the numbers in all rows, columns and the
two diagonal are the same.  It can therefore be modeled in CSP by
considering $N^2$ variables with initial domains $\{1, 2, \ldots
,N^2\}$ together with linear equation constraints and a global
\emph{all-different} constraint stating that all variables should have
a different value. The constant value that should be the sum of all
rows, columns and the two diagonals can be easily computed to be
$N(N^2 +1)/2$.

\subsubsection{All-Interval Problem (AIP)}
The All-Interval Problem (\texttt{prob007} in CSPLib) consists of
composing a sequence of $N$ notes such that all are different and tonal
intervals between consecutive notes are also distinct. This problem is
equivalent to finding a permutation of the $N$ first integers such that the
absolute difference between two consecutive pairs of numbers are all
different. This amounts to finding a permutation $(X_1, \ldots ,X_N)$ of $(0,
\ldots ,N-1)$ such that the list $(abs(X_1-X_2),abs(X_2-X_3), \ldots
,abs(X_{N-1} - X_N))$ is a permutation of $(1, \ldots ,N-1)$. A possible
solution for $N = 8$ is $(3, 6, 0, 7, 2, 4, 5, 1)$ because all consecutive
distances are different.

\subsubsection{Number Partitioning Problem (NPP)}
The Number Partitioning Problem (\texttt{prob049} in CSPLib) consists of
finding a partition of numbers $\{1, \ldots ,N\}$ into two groups A
and B of the same cardinality such that the sum of numbers in A is
equal to the sum of numbers in B and the sum of squares of numbers in
A is equal to the sum of squares of numbers in B. A solution for $N =
8$ is $A = \{1, 4, 6, 7\}$ and $B = \{ 2, 3, 5, 8\}$.

\subsubsection{Costas Array Problem (CAP)}
The Costas Array Problem consists of filling an $N \times N$ grid with $N$ marks such
that there is exactly one mark per row and per column and the $N(N-1)/2$
vectors joining the marks are all different. It is convenient to see the
Costas Array Problem as a permutation problem by considering an array of $N$
variables $(X_1, \ldots , X_n)$ which forms a permutation of $\{1, 2, \ldots
, N\}$ subject to some \emph{all-different} constraints (see~\cite{diaz12} for a
detailed modeling). This problem has many practical applications and
currently it has a whole community active working around it
(\url{http://www.costasarrays.org/}).

\section{Performance Analysis}
\label{sec:PerformanceAnalysis}
This section presents and discusses our experimental results of the
X10 implementation of the Adaptive Search algorithm.

\ifx{1}{2}
\subsection{Results: Functional parallelism X10 implementation}
Table~\ref{tab:FP1Result} shows the results of the \emph{first
  version} of the \emph{functional parallelism} X10
implementation. Only two benchmarks (2 instances of MSP and CAP) are
presented. Indeed, we did not investigate much more this approach
since the results are clearly not good.  Each problem instance was
executed with a variable number of activities (\textit{THNUM = 2, 4
  and 8}). Is worth noticing, that the environmental X10 variable
\emph{X10\_NTHREADS} was passed to the program with an appropriate
value to each execution. This variable controls the number of initial
working threads per place in the X10 runtime system.

\begin{table}[htb]
  \begin{center}
  \begin{tabular}{||c|c|c|c|c|c||}
    \hline
    ~Problem~ & time (s) & \multicolumn{3}{|c|}{speed-up with k activ.} & time (s) \\
    \cline{3-5}
    instance  & seq.     & ~~~~2~~~~ & ~~~~4~~~~ & ~~~~8~~~~            & 8 activ.\\
    \hline
    \hline
    \texttt{MSP-100} & 11.98 & 0.86 & 0.95 & 0.77 & 15.49 \\
    \texttt{MSP-120} & 24.17 & 1.04 & 0.97 & 0.98 & 24.65 \\
    \hline
    \texttt{CAP-17}  & ~1.56 & 0.43 & 0.28 & 0.24 & ~6.53 \\
    \texttt{CAP-18}  & 12.84 & 0.51 & 0.45 & 0.22 & 57.16 \\
    \hline
  \end{tabular}
  \end{center}
  \caption{Functional parallelism -- first approach (timings and speed-ups)}
  \label{tab:FP1Result}
\end{table}

As seen in Table~\ref{tab:FP1Result}, for all the treated cases the
obtained speed-up is less than $1$ (i.e. a slowdown factor), showing a
deterioration of the execution time due to this parallel
implementation. So, it is possible to conclude that no gain time is
obtainable in this approach. To analyze this behavior it is important
to return to the description of the Listing~\ref{fig:as-x10FP1}.  As
already noted, the parallel function \emph{selVarHighCost} in this
implementation are located into the main loop of the algorithm, so
$THNUM$ activities are created, scheduled and synchronized at each
iteration in the program execution, being a very important source of
overhead.  The results we obtained suggest that this overhead is
larger than the improvement obtained by the implementation of this
parallel strategy.

Turning to the \emph{second approach}, Table~\ref{tab:FP2Result} shows
the results obtained with this strategy. Equally, the number of
activities spawn, in this case at the beginning, was varied from 2 to
8.

\begin{table}[htb]
  \begin{center}
  \begin{tabular}{||c|c|c|c|c|c||}
    \hline
    ~Problem~ & time (s) & \multicolumn{3}{|c|}{speed-up with k activ.} & time (s) \\
    \cline{3-5}
    instance  & seq.     & ~~~~2~~~~ & ~~~~4~~~~ & ~~~~8~~~~            & 8 activ.\\
    \hline
    \hline
    \texttt{MSP-100} & 11.98 & 1.15 & 0.80 & 0.86 & 13.87 \\
    \texttt{MSP-120} & 24.17 & 1.23 & 0.94 & 0.63 & 38.34 \\
    \hline
    \texttt{CAP-17}  & ~1.56 & 0.56 & 0.30 & 0.25 & ~6.35 \\
    \texttt{CAP-18}  & 12.84 & 0.74 & 0.39 & 0.27 & 46.84 \\
    \hline
  \end{tabular}
  \end{center}
  \caption{Functional parallelism -- second approach (timings and speed-ups)}
  \label{tab:FP2Result}
\end{table}

Even if the results are slightly better, there is no noticeable
speed-up. This is due to a new form of overhead due to the synchronization
mechanism which is used in the inner loop of the algorithm to assign tasks
and to wait for their termination (see Listing~\ref{fig:as-x10FP2}).

\subsection{Results: Data Parallel X10 Implementation}
\fi{}

We present the experimental results of the Adaptive Search algorithm
in its X10 data parallel implementation.  We do not present results
for functional parallelism because the tests we carried out show that
the granularity of individual threads is too fine to yield any
performance improvement.

The testing environment used in each running was a non-uniform memory
access (NUMA) computer, with 2 Intel Xeon W5580 CPUs each one with 4
hyper-threaded cores running at 3.2GHz. This system has 12 GB of main
memory.

At the software level, the X10 runtime system can be deployed in two
different backends: Java backend and \texttt{C++} backend; they differ
in the native language used to implement the X10 program (Java or
\texttt{C++}), also they present different trade-offs on different
machines. Currently, the \texttt{C++} backend seems relatively mature
and faster, therefore, we have chosen it for this experimentation.

Regarding the stochastic nature of the Adaptive Search behavior,
several executions of the same problem were done and the times
averaged.  We ran 100 samples for each experimental case in the
benchmark.

In this presentation, all tables use the same format: the first column
identifies the problem instance, the second column is the execution
time of the problem in the sequential implementation, the next group
of columns contains the corresponding speed-up obtained with a varying
number of cores (activities or places), and the last column presents
the execution time of the problem with the highest number of cores.

\BL \noindent \textbf{Magic Square
  Problem}. Table~\ref{tab:MagicSquareResult} presents the data
obtained solving several large instances of MSP. Both raw times
(average of 100 runs) and relative speed-ups are reported. The results
show quasi-linear speed-ups (which seem independent of the size of the
problem).

\begin{table}[htb]
  \begin{center}
  \begin{tabular}{||c|c|c|c|c|c|c||}
    \hline
    ~Problem~ & time (s) & \multicolumn{4}{|c|}{speed-up with k places}  & time (s) \\
    \cline{3-6}
    instance  & seq.     & ~~~~2~~~~ & ~~~~4~~~~ & ~~~~6~~~~ & ~~~~8~~~~ &~8 places~\\
    \hline
    \hline
    \texttt{40} & ~~0.47 & 2.12 & 3.20 & 4.14 & 4.42 & 0.11 \\
    \texttt{60} & ~~1.59 & 1.58 & 2.68 & 3.25 & 3.62 & 0.44 \\
    \texttt{80} & ~~5.10 & 1.84 & 2.92 & 3.72 & 4.19 & 1.22 \\
    \texttt{100}&  11.88 & 1.69 & 2.92 & 3.64 & 4.36 & 2.72 \\
    \hline
  \end{tabular}
  \end{center}
  \caption{Magic Square: data parallel (timings and speed-ups)}
  \label{tab:MagicSquareResult}
\end{table}

\BL \noindent \textbf{All-Interval
  Problem}. Table~\ref{tab:AllIntervalResult} shows the average time
of several instances of AIP together with the speed-ups acquired with
different number of places. Here again the speed-ups are practically
linear up to 5.38 with 8 places. Moreover, in this case the speed-up
tends to increase with the size of the problem.

\begin{table}[htb]
  \begin{center}
  \begin{tabular}{||c|c|c|c|c|c|c||}
    \hline
    ~Problem~ & time (s) & \multicolumn{4}{|c|}{speed-up with k places}  & time (s) \\
    \cline{3-6}
    instance  & seq.     & ~~~~2~~~~ & ~~~~4~~~~ & ~~~~6~~~~ & ~~~~8~~~~ &~8 places~\\
    \hline
    \hline
    \texttt{50}  & 0.027 & 2.25 & 3.24 & 4.46 & 4.94 & 0.005 \\
    \texttt{100} & 0.47~ & 2.12 & 3.21 & 4.32 & 4.96 & 0.09~ \\
    \texttt{150} & 2.36~ & 1.74 & 3.47 & 4.65 & 5.15 & 0.46~ \\
    \texttt{200} & 8.29~ & 1.84 & 3.66 & 5.28 & 5.38 & 1.54~ \\
    \hline
  \end{tabular}
  \end{center}
  \caption{All-interval: data parallel (timings and speed-ups)}
  \label{tab:AllIntervalResult}
\end{table}

\BL \noindent \textbf{Number Partitioning
  Problem}. Table~\ref{tab:PartitResult} shows the results of NPP.
From this data, it can be seen that the corresponding speed-up
obtained is almost the ideal speed-up for each number of places used.
Also, the speed-up increases linearly with the number of cores
to reach a maximum of 7.64 with 8 places.  Moreover, this speed-up
seems to increase with the size of the problem.

\begin{table}[htb]
  \begin{center}
  \begin{tabular}{||c|c|c|c|c|c|c||}
    \hline
    ~Problem~ & time (s) & \multicolumn{4}{|c|}{speed-up with k places}  & time (s) \\
    \cline{3-6}
    instance  & seq.     & ~~~~2~~~~ & ~~~~4~~~~ & ~~~~6~~~~ & ~~~~8~~~~ &~8 places~\\
    \hline
    \hline
    \texttt{1400} & 1.07 & 1.50 & 2.60 & 4.49 & 5.00 & 0.21 \\
    \texttt{1600} & 1.94 & 1.78 & 3.59 & 5.10 & 6.97 & 0.28 \\
    \texttt{1800} & 2.30 & 1.62 & 3.48 & 3.98 & 5.49 & 0.42 \\
    \texttt{2000} & 4.31 & 2.34 & 4.87 & 7.06 & 7.64 & 0.56 \\
    \hline
  \end{tabular}
  \end{center}
  \caption{Partition: data parallel (timings and speed-ups)}
  \label{tab:PartitResult}
\end{table}

\BL \noindent \textbf{Costas Array Problem}.
Table~\ref{tab:CAPResult} presents the average time for several instances of
the Costas Array Problem together with the speed-up obtained when using
different numbers of places. The data confirm the trends above observed.
Note that the best speed-up (9.56) is super-linear and is obtained for the most difficult
instance of CAP (size of 19): the execution time is drastically reduced from
103.98 seconds to only 10.87 seconds on 8 places.

\begin{table}[htb]
  \begin{center}
  \begin{tabular}{||c|c|c|c|c|c|c||}
    \hline
    ~Problem~ & time (s) & \multicolumn{4}{|c|}{speed-up with k places}  & time (s) \\
    \cline{3-6}
    instance  & seq.     & ~~~~2~~~~ & ~~~~4~~~~ & ~~~~6~~~~ & ~~~~8~~~~ &~8 places~\\
    \hline
    \hline
    \texttt{16} & ~~~0.26 & 1.79 & 3.52 & 4.00 & 7.09 & ~~0.04 \\
    \texttt{17} & ~~~1.94 & 1.96 & 4.46 & 6.47 & 9.84 & ~~0.19 \\
    \texttt{18} & ~~10.90 & 1.68 & 4.05 & 5.09 & 6.88 & ~~1.59 \\
    \texttt{19} & 103.98 & 2.13 & 4.63 & 5.81 & 9.56 & 10.87 \\
    \hline
  \end{tabular}
  \end{center}
  \caption{Costas Array: data parallel (timings and speed-ups)}
  \label{tab:CAPResult}
\end{table}

\begin{figure}[htb]
\centerline{\fbox{{\includegraphics[scale=0.55]{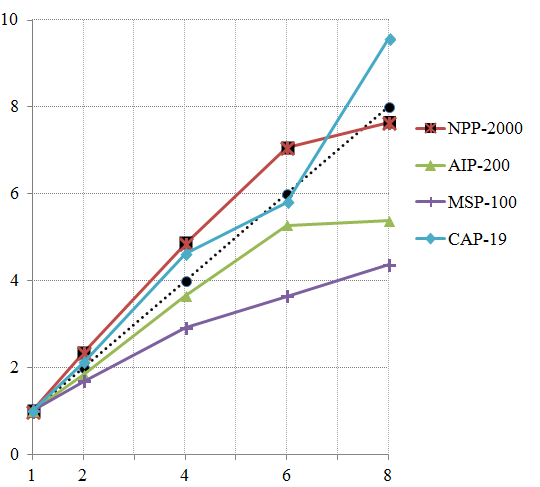}}}} 
\caption{Speed-ups for the most difficult instance of each problem}\label{fig:BenchmarkSpeed-Ups}
\end{figure}

Figure~\ref{fig:BenchmarkSpeed-Ups} shows the
speed-ups reached on the most difficult instance of each problem. It can be
seen that the speed-up increases almost linearly with the number of
places used in the X10 program.

The performance evaluation developed in this work shows that a
parallel Local Search solver implemented in X10 has good
performance\footnote{Within a bound of 4-5 w.r.t.~our sequential C
  implementation.}  using a data parallel strategy.  The study has
identified constant behavior of the speed-up with relation to the size
of the problem (for some problems the speed-up improves with the size
of the problem).  The resulting average runtime and the speed-ups
obtained in the entire experimental test performed are as good as
reported in the literature when using other IPC frameworks such as
MPI~\cite{Diaz2011,diaz12,DBLP:conf/evoW/CaniouCDA11} and seems to lie
within the predictable bounds proposed
by~\cite{DBLP:journals/corr/abs-1212-4287}.


\section{Conclusion and Future Work}

We presented a parallel X10 implementation of an effective Local
Search algorithm, Adaptive Search.  We first experimented with
functional parallelism, i.e. trying to divide the inner loop of the
algorithm into various concurrent tasks.  As expected, this yielded no
speed-up, mainly because of the bookkeeping overhead (creation,
scheduling and synchronization) that are too fine-grained.

We then proceeded with a data parallel implementation, in which the
search space is decomposed in possible different random initial
configurations of the problem and getting isolated solver instances to
work on each point concurrently.  We got a good level of performance
for the X10 data-parallel implementation, reaching a maximum speed-up
of 9.84 with 8 \emph{places} for the Costas Array Problem.  Linear (or
close) speed-ups have been recorded in all problems we studied and
they remain constant (or increasing) wrt the size of the problem.

X10 has proved a suitable platform to exploit parallelism in different
ways for constraint-based local search solvers, ranging from single
shared memory inter-process parallelism to more external distributed
memory programming model.  Additionally, the use of the X10 implicit
communication mechanisms shows that X10 enables one to abstract the
complexity of the parallelism with a very simple model, e.g.~the
distributed arrays and the termination detection system in our data
parallel implementation.


Future work will focus on the implementation of a cooperative Local
Search parallel solver using data parallelism.  The key idea is to
take advantage of all communications tools available in this APGAS
model, to exchange information between different solver instances in
order to obtain a more efficient and scalable solver implementation.
We also plan to test the behavior of a cooperative implementation,
under different HPC architectures, such as the many-core Xeon PHI,
GPGPU accelerators and grid computing platforms like Grid5000.

\bibliographystyle{plain}
\bibliography{library}

\end{document}